# UNA VALUTAZIONE DI COPERTURA, QUALITÀ ED EFFICIENZA DEI SERVIZI SANITARI REGIONALI TRA 2010 – 2013

di Filippo Elba[1]

**Breve introduzione metodologica**

Nelle pagine che seguono è proposto sia un utilizzo abituale, sia un utilizzo finalizzato alla determinazione di una "best performance" della tecnica Data Envelopment Analysis (DEA) su cui si basa il modello Multiplicative Non-Parametric Corporate Performance (MNCP) di Emrouznejad e Cabanda (2010)[2]. DEA, infatti, è abitualmente utilizzata con il fine di individuare una frontiera di produzione efficiente, tuttavia è frequente l'utilizzo con finalità di benchmarking, come ben evidenziato da Cook et al. (2004): "Although DEA has a strong link to production theory in economics, the tool is also used for benchmarking in operations management [...] In the circumstance of benchmarking, the efficient DMUs, as defined by DEA, may not necessarily form a production frontier, but rather lead to a best-practice frontier". In quest'ottica, DEA è qui utilizzata per sintetizzare in un unico indice indicatori diversi che, presi singolarmente, sono utili a descrivere solo aspetti parziali (nello specifico la copertura di alcuni servizi sanitari). Per la realizzazione di tale indice sintetico è perciò sfruttato il meccanismo di ponderazione insito in DEA il quale mette ogni unità analizzata nelle condizioni ottimali, attribuendo ad ognuna di esse la possibilità di far pesare maggiormente gli indicatori migliori. In tal modo, in assenza di pesi esogeni, ogni unità è messa nella condizione di potersi esprimere al meglio.

**1. La spesa sanitaria in Italia**

Negli ultimi anni le risorse destinate alla sanità, in Italia, sono oggetto di tagli. La spesa complessiva, sempre cresciuta tra 2008 e 2011, seppur in maniera via via più contenuta (tra 2010 e 2012 è rimasta sostanzialmente ferma a 112,7 mld), subisce una riduzione di un mld di € tra 2012 e 2013, per poi riportarsi a 112,7 mld nel 2014 (Tab. 1.1).
La situazione non è, tuttavia, condivisa alla stessa maniera da tutte le regioni. Tra quelle in piano di rientro, Lazio, Campania e Calabria fronteggiano tagli già a partire dal 2009, rispetto a regioni come la Lombardia che, al contrario, vede crescere la sua spesa lungo tutto il periodo considerato, passando da 16,7 mld di € del 2008 a 18,8 mld di € del 2014 (Tab. 1.1).

---

[1] Assistente di insegnamento e ricerca presso il Dipartimento di Scienze per l'Economia e l'Impresa dell'Università degli Studi di Firenze (filippo.elba@unifi.it).
[2] Il presente lavoro è un'estensione di Elba et al. (2017).

**Tab. 1.1: spesa sanitaria corrente, anni 2008-2014**

*Valori in migliaia di euro*

| | 2008 | 2009 | 2010 | 2011 | 2012 | 2013 | 2014 |
|---|---|---|---|---|---|---|---|
| PIEMONTE | 8.168.765 | 8.444.150 | 8.576.644 | 8.534.730 | 8.454.187 | 8.268.504 | 8.257.614 |
| VALLE D'AOSTA | 260.879 | 264.043 | 278.060 | 279.486 | 279.298 | 272.649 | 260.785 |
| LOMBARDIA | 16.740.240 | 17.222.431 | 17.844.158 | 18.186.558 | 18.307.208 | 18.446.212 | 18.870.104 |
| PROV AUT BOLZANO | 1.108.183 | 1.065.860 | 1.099.606 | 1.112.280 | 1.158.176 | 1.160.304 | 1.141.812 |
| PROV AUT TRENTO | 995.402 | 1.062.713 | 1.095.332 | 1.131.258 | 1.157.508 | 1.150.646 | 1.153.187 |
| VENETO | 8.652.843 | 8.907.065 | 9.050.809 | 9.019.565 | 8.915.056 | 8.783.767 | 8.788.140 |
| FRIULI VENEZIA GIULIA | 2.316.504 | 2.414.483 | 2.448.022 | 2.500.544 | 2.498.732 | 2.475.931 | 2.385.568 |
| LIGURIA | 3.186.542 | 3.288.245 | 3.273.850 | 3.257.367 | 3.171.880 | 3.135.759 | 3.168.488 |
| EMILIA ROMAGNA | 8.061.983 | 8.461.042 | 8.631.331 | 8.731.365 | 8.892.326 | 8.744.020 | 8.768.698 |
| TOSCANA | 6.805.062 | 7.261.649 | 7.252.161 | 7.255.084 | 7.282.498 | 7.131.197 | 7.260.237 |
| UMBRIA | 1.567.200 | 1.623.697 | 1.636.560 | 1.647.572 | 1.658.570 | 1.650.583 | 1.645.404 |
| MARCHE | 2.647.113 | 2.761.392 | 2.835.884 | 2.837.582 | 2.786.463 | 2.743.929 | 2.768.653 |
| LAZIO | 10.987.886 | 11.250.297 | 11.143.017 | 11.007.694 | 10.925.879 | 10.701.872 | 10.682.689 |
| ABRUZZO | 2.371.856 | 2.362.669 | 2.355.425 | 2.323.540 | 2.330.702 | 2.290.309 | 2.348.869 |
| MOLISE | 653.501 | 668.710 | 665.753 | 654.242 | 663.517 | 652.881 | 661.353 |
| CAMPANIA | 10.084.763 | 10.246.143 | 10.116.740 | 9.949.429 | 9.674.527 | 9.510.214 | 9.716.016 |
| PUGLIA | 7.131.501 | 7.197.239 | 7.289.302 | 7.131.526 | 7.029.857 | 7.022.325 | 7.151.521 |
| BASILICATA | 1.020.474 | 1.042.555 | 1.063.916 | 1.068.467 | 1.038.555 | 1.028.247 | 1.037.458 |
| CALABRIA | 3.384.485 | 3.514.678 | 3.473.073 | 3.400.489 | 3.332.621 | 3.297.242 | 3.360.408 |
| SICILIA | 8.341.115 | 8.471.903 | 8.606.583 | 8.610.897 | 8.546.649 | 8.566.884 | 8.579.301 |
| SARDEGNA | 2.944.030 | 3.082.855 | 3.165.745 | 3.217.523 | 3.262.632 | 3.233.326 | 3.266.518 |
| **ITALIA** | **108.143.924** | **111.372.503** | **112.630.340** | **112.623.941** | **112.688.218** | **111.684.110** | **112.672.629** |
| regioni non in piano di rientro | 48.681.457 | 50.568.076 | 51.588.669 | 52.003.560 | 52.052.556 | 51.663.714 | 52.307.182 |
| regioni in piano di rientro e commissariate | 27.482.491 | 28.042.497 | 27.754.008 | 27.335.394 | 26.927.246 | 26.452.518 | 26.769.335 |
| regioni in piano di rientro | 23.641.381 | 24.113.292 | 24.472.529 | 24.277.153 | 24.030.693 | 23.857.713 | 23.988.436 |
| regioni a statuto speciale | 7.624.998 | 7.889.954 | 8.086.765 | 8.241.091 | 8.356.346 | 8.292.856 | 8.207.870 |

*Fonte: elaborazione Agenas su dati modelli Ce consuntivi 2008-2014 (NSIS)*

A livello UE, la situazione italiana è in lieve controtendenza rispetto a quella dei principali partner. Per quanto riguarda, infatti, la quota di spesa sanitaria sul Pil, l'Italia passa dal 7,5% del 2009, al 7,1% 2015. Nello stesso periodo, la Francia passa dal 7,9% all'8,2%, la Germania dal 7,1% al 7,2% e il Regno Unito dal 7,8% al 7,6%. In termini di percentuale di spesa sanitaria sul totale della spesa pubblica, l'Italia passa dal 14,6% del 2009 al 14,1% del 2015, La Francia dal 14% al 14,3%, la Germania dal 15 al 16,3% e il Regno Unito dal 16,2% al 17,8% (Fonte Eurostat[3]).

Nelle pagine che seguono è proposta una valutazione riferita a grado di copertura dei servizi sanitari regionali e della loro relativa efficienza. Innanzitutto viene analizzata copertura e qualità dei servizi ospedalieri (Sez. 2), quindi quella dei servizi distrettuali, ossia tutti quei servizi di natura sanitaria che vengono erogati al di fuori delle strutture ospedaliere[4] (Sez. 3). Si conclude con una valutazione dell'efficienza con cui vengono garantiti questi servizi, intesa come capacità di assicurare l'attuale livello di prestazioni (al di là dell'essere ottimale o meno) al più basso costo possibile (Sez. 4).

---

[3] http://appsso.eurostat.ec.europa.eu/nui/show.do?dataset=gov_10a_exp&lang=en
[4] In realtà, una parte delle prestazioni sanitarie fornite al di fuori dell'ambito ospedaliero costituisce quella che viene definita "assistenza sanitaria collettiva in ambiente di vita e di lavoro". Essendo il ruolo di questa residuale (in termini di costo, pesa circa il 4% dell'intera spesa nazionale sanitaria), non viene presa in considerazione nel presente report.

## 2. La valutazione dei servizi ospedalieri regionali: 2010 – 2013

La spesa per servizi ospedalieri rappresenta all'incirca il 45% della spesa sanitaria complessiva (fatte le dovute distinzioni tra regioni). Partendo dalle informazioni disponibili nella banca dati Istat Health For All (HFA)[5], è qui proposta una valutazione dei servizi ospedalieri regionali nel periodo 2010 – 2013. L'analisi si compone, da un lato della valutazione della copertura dei servizi ospedalieri rispetto alla popolazione di riferimento, dall'altro della valutazione della qualità degli stessi, in base ai giudizi dei ricoverati[6].

Con riferimento alla costruzione dell'Indice di Copertura del Servizio Ospedaliero (ICSO), le variabili prese in considerazione sono:

- tasso posti letto ospedalieri day hospital (per 10 mila abitanti);
- tasso posti letto ospedalieri ordinari (per 10 mila abitanti);
- tasso medici, odontoiatri istituti cura pubblici, privati accreditati (per 10 mila abitanti);
- tasso personale infermieristico istituti cura pubblici, privati accreditati (per 10 mila abitanti);
- tasso personale funzioni riabilitazione istituti cura pubblici, privati accreditati (per 10 mila abitanti);
- tasso personale tecnico-sanitario istituti cura pubblici, privati accreditati (per 10 mila abitanti).

Partendo da questi tassi, la sintesi, con costruzione di un unico indice, è realizzata ricorrendo al modello MNCP basato su DEA.

Per quel che concerne, invece, la costruzione dell'Indice di Qualità del Servizio Ospedaliero (IQSO), le variabili prese in considerazione sono:

- persone molto soddisfatte per assistenza medica ospedaliera (in percentuale dei ricoverati);
- persone molto soddisfatte assistenza infermieristica ospedaliera (in percentuale dei ricoverati);
- persone molto soddisfatte servizi igienici ospedalieri (in percentuale dei ricoverati).

In questo caso, anche per omogeneità rispetto alle grandezze misurate, la sintesi in un unico indicatore è realizzata attraverso media semplice[7].

### 2.1 L'Indice di Copertura del Servizio Ospedaliero

L'ICSO si configura come un indice di sintesi riferito alla capacità regionale di erogare determinati servizi ospedalieri (vedi lista paragrafo precedente) in rapporto alla popolazione locale. Il valore dell'indice è compreso tra 0 (valore minimo) e 1 (copertura massima).

Guardando al valore medio annuo degli ICSO regionali, si evidenzia come, se tra il 2010 ed il 2011 questo si incrementa, passando da 0,64 a 0,68, ciò non è più vero tra il 2011 e gli anni successivi. Addirittura, nel 2013, il valore dell'ICSO medio delle regioni italiane è pari a 0,6, più basso di quanto rilevato quattro anni prima (Fig. 2.1).

Si sottolinea che, essendo rimasto sostanzialmente uguale il valore della deviazione standard nel periodo (oscillando tra lo 0,24 e lo 0,26), la variabilità dei punteggi ottenuti dalle singole regioni è rimasto, in media, costante nel tempo.

---

[5] http://www.istat.it/it/archivio/14562

[6] La valutazione qui proposta ha un obbiettivo differente rispetto a quello, per esempio, del monitoraggio di Livelli Essenziali Assistenziali (o LEA). Nel caso dei LEA, infatti, l'intento è quello di valutare se le regioni sono in grado di garantire un livello minimo, ritenuto essenziale, di servizi. Nella valutazione qui proposta, invece, l'intento è quello di individuare le migliori performance regionali in materia di servizi ospedalieri, in modo da individuare quella che viene generalmente indicata come una best practice, che diventa punto di riferimento per misurare la "distanza" delle regioni con performance scarse.

[7] Si rinvia a lavori futuri l'eventualità di sostituire la media semplice con una media ponderata sulla base di indicazioni da parte del policy maker o di un esperto.

**Fig. 2.1: ICSO nel periodo 2010 – 2013: media valori regionali per anno**

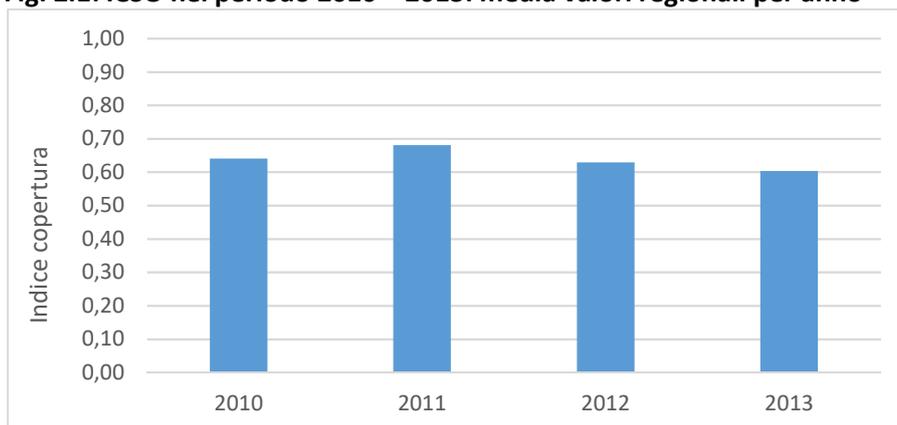

Andando a guardare la situazione dettagliata per singola regione, emergono interessanti differenze. Le regioni che, abbastanza stabilmente, mantengono un ICSO elevato o massimo durante tutto il periodo sono: l'Alto Adige, l'Emilia Romagna, il Friuli Venezia Giulia, la Liguria (Tab. 2.1).

Tra le regioni che, al contrario, restano agli ultimissimi posti durante tutto il periodo ci sono Sicilia e Campania (Tab. 2.1).

Le regioni che, nel tempo, migliorano la propria performance sono la Valle d'Aosta, il Trentino e, in maniera più limitata, Basilicata e Veneto (Tab. 2.1).

La Lombardia è la regione che, invece, peggiora in maniera più evidente, in particolare tra 2011 e anni successivi (Tab. 2.1).

In generale, tra le regioni che primeggiano vi sono quelle del Nord Italia, soprattutto le meno estese e a statuto speciale. Tra le regioni con l'ICSO più basso, ci sono per lo più quelle meridionali (Tab. 2.1).

**Tab. 2.1: ICSO regionali nel periodo 2010 - 2013**

| Regione | 2010 | 2011 | 2012 | 2013 |
|---|---|---|---|---|
| Abruzzo | 0,533 | 0,506 | 0,548 | 0,495 |
| Alto Adige | 1,000 | 0,888 | 1,000 | 1,000 |
| Basilicata | 0,585 | 0,577 | 0,587 | 0,769 |
| Calabria | 0,391 | 1,000 | 0,656 | 0,554 |
| Campania | 0,317 | 0,282 | 0,285 | 0,247 |
| Emilia-Romagna | 1,000 | 1,000 | 0,925 | 0,828 |
| Friuli-Venezia Giulia | 1,000 | 1,000 | 1,000 | 0,986 |
| Lazio | 1,000 | 1,000 | 0,784 | 0,698 |
| Liguria | 0,795 | 1,000 | 1,000 | 1,000 |
| Lombardia | 0,798 | 0,799 | 0,275 | 0,290 |
| Marche | 0,487 | 0,697 | 0,414 | 0,371 |
| Molise | 1,000 | 0,505 | 0,615 | 0,473 |
| Piemonte | 0,769 | 0,815 | 0,838 | 0,790 |
| Puglia | 0,533 | 0,485 | 0,325 | 0,281 |
| Sardegna | 0,353 | 0,412 | 0,284 | 0,263 |
| Sicilia | 0,251 | 0,219 | 0,262 | 0,260 |
| Toscana | 0,492 | 0,530 | 0,534 | 0,523 |
| Trentino | 0,673 | 0,805 | 0,867 | 0,790 |
| Umbria | 0,380 | 0,500 | 0,478 | 0,462 |
| Valle d'Aosta | 0,619 | 0,721 | 1,000 | 1,000 |
| Veneto | 0,487 | 0,558 | 0,545 | 0,597 |

**Legenda:** la scala cromatica utilizzata va dal rosso intenso (ICSO basso) al verde intenso (ICSO alto).

## 2.2 L'indice di Qualità del Servizio Ospedaliero

Avendo visto le differenze cronologiche e territoriali riguardanti la quantità di servizi ospedalieri offerti rispetto alla popolazione, utile è guardare anche a quello che succede da un punto di vista più prettamente qualitativo. L'IQSO sintetizza in un unico indice la quota di pazienti molto soddisfatti dai servizi ospedalieri usufruiti (vedi lista primo paragrafo). Il suo valore è compreso tra 0 (qualità scarsa) a 100 (qualità massima). Il livello medio regionale della qualità dei servizi ospedalieri è sostanzialmente costante lungo l'arco di tempo considerato. Il valore oscilla tra il 38 ed il 40% (Fig. 2.2).
Deviazione standard limitata e, anche in questo caso, constante negli anni, a segnalare come la variabilità dei punteggi ottenuti dalle singole regioni sia rimasto, in media, costante nel tempo (valori compresi tra 12 e 15%).

**Fig. 2.2: IQSO nel periodo 2010 – 2013: media valori regionali per anno**

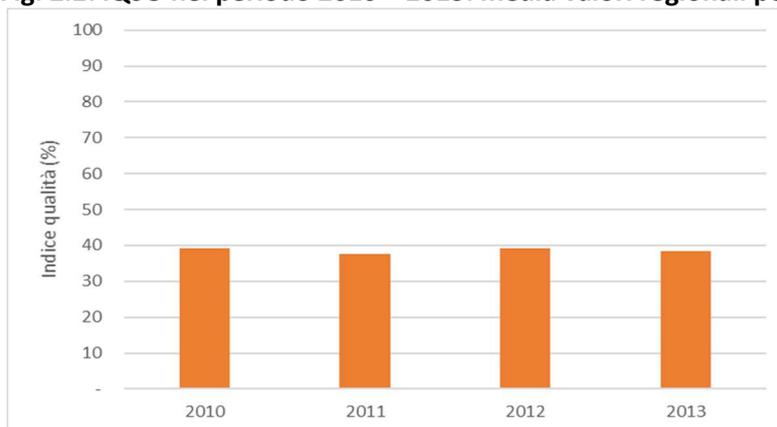

Per quanto concerne il dettaglio regionale, si evidenzia come le regioni che durante l'arco di tempo considerato registrano la quota più elevati di pazienti molto soddisfatti dei servizi ospedalieri sono: il Veneto, l'Umbria, il Piemonte, l'Emilia Romagna, il Friuli Venezia Giulia e le due provincie autonome (Tab. 2.2).
Tra le regioni che presentano l'IQSO più basso Sicilia, Puglia, Campania e Calabria (Tab. 2.2).
Il Molise peggiora la performance nel tempo, mentre il Lazio la migliora (Tab. 2.2).

**Tab. 2.2: IQS regionali nel periodo 2010 – 2013**

| Regione | 2010 | 2011 | 2012 | 2013 |
|---|---|---|---|---|
| Abruzzo | 27,670 | 26,410 | 24,033 | 28,750 |
| Alto Adige | 56,330 | 54,253 | 66,550 | 59,460 |
| Basilicata | 28,323 | 33,070 | 28,413 | 24,863 |
| Calabria | 23,050 | 16,883 | 27,233 | 29,453 |
| Campania | 15,590 | 23,030 | 17,167 | 24,237 |
| Emilia-Romagna | 43,430 | 62,730 | 62,793 | 50,437 |
| Friuli-Venezia Giulia | 50,690 | 44,653 | 63,797 | 41,040 |
| Lazio | 25,953 | 30,417 | 36,193 | 38,217 |
| Liguria | 42,480 | 30,443 | 43,367 | 42,723 |
| Lombardia | 45,363 | 43,223 | 51,277 | 47,907 |
| Marche | 28,760 | 34,993 | 29,500 | 40,673 |
| Molise | 31,437 | 32,847 | 32,313 | 17,683 |
| Piemonte | 51,930 | 42,560 | 46,893 | 50,430 |
| Puglia | 20,310 | 28,557 | 20,497 | 23,997 |
| Sardegna | 34,527 | 33,363 | 41,360 | 26,170 |
| Sicilia | 21,487 | 18,507 | 16,943 | 15,113 |
| Toscana | 40,180 | 36,243 | 35,797 | 34,980 |
| Trentino | 75,267 | 61,457 | 59,257 | 66,763 |
| Umbria | 45,670 | 42,743 | 37,650 | 50,977 |
| Valle d'Aosta | 61,980 | 51,780 | 26,967 | 37,973 |
| Veneto | 54,647 | 42,043 | 56,947 | 52,783 |

**Legenda:** la scala cromatica utilizzata va dal rosso intenso (ICSO basso) al verde intenso (ICSO alto).

**2.3 Copertura e Qualità**

Dopo aver guardato i due indicatori separatamente, si procede incrociando le informazioni, in modo da poter valutare se e per quali regioni un ICSO elevato si accompagna ad un IQSO soddisfacente e viceversa. Per far questo ci si avvale di analisi grafica. Grazie ad essa è possibile suddividere il piano cartesiano in quattro quadranti e le regioni in altrettanti gruppi[8]:

- quadrante in alto a destra – "Alta copertura, Alta qualità": è l'area in cui si trovano le regioni che ottengono le migliori performance sia dal punto di vista della capacità di copertura del servizio ospedaliero, sia in termini di qualità del servizio offerto;
- quadrante in basso a destra – "Bassa copertura, Alta qualità": è l'area in cui si trovano le regioni che, pur ottenendo buone performance dal punto di vista della qualità del servizio offerto, non sono in grado di garantire livelli di copertura paragonabili a quelli delle regioni presenti nel quadrante superiore;
- quadrante in basso a sinistra – "Bassa copertura, Bassa qualità": è l'area in cui si trovano le regioni che fanno peggio in termini di servizi ospedalieri, avendo un basso ICSO ed un altrettanto basso IQSO;
- quadrante in alto a sinistra – "Alta copertura, Bassa qualità": è l'area in cui si trovano le regioni che, seppur in grado di offrire una buona copertura di servizi ospedalieri rispetto alla popolazione, non eccellono per quel che riguarda la qualità del servizio offerto.

Tra le regioni di eccellenza nel 2010 ci sono le province autonome, le regioni a statuto speciale del Nord, la Liguria, il Piemonte, la Lombardia, L'Emilia Romagna (Fig. 2.3).

Tra le regioni con le performance peggiori ci sono tutte quelle del Meridione, fatta eccezione per il Molise che, pur avendo un ICSO elevato, si contraddistingue per un livello dell'IQSO medio basso, e le Marche (Fig. 2.3).

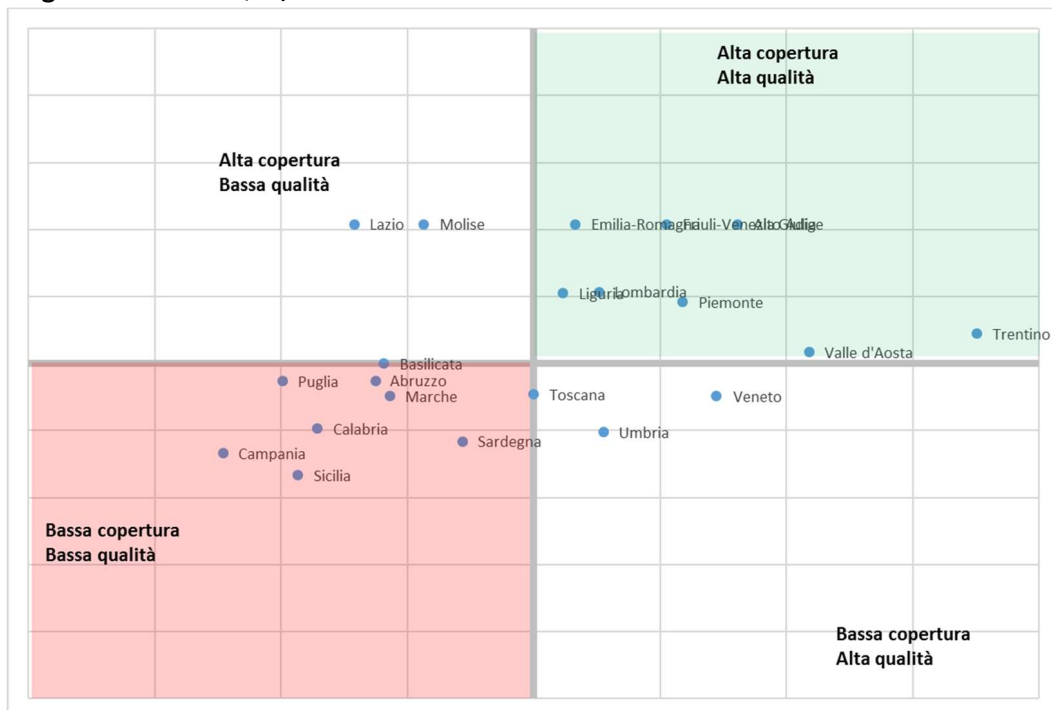

**Fig. 2.3: ICSO vs IQSO, 2010**

---

[8] Gli assi delle ascisse e delle ordinate vengono fatti passare in corrispondenza dei valori mediani di ICSO e IQSO, in modo tale da "spaccare" i campioni in due, incrociando poi i risultati.

La situazione del 2011 è pressappoco la medesima dell'anno precedente. Tra le principali differenze, il fatto che Liguria e Calabria si spostano nel quadrante "Alta copertura, Bassa Qualità", pur venendo da situazioni diametralmente opposte. Oltre a ciò, il Molise raggiunge le altre regioni meridionali del quadrante "Bassa copertura, Bassa qualità" e le Marche assumono il valore mediano sia per quel che concerne l'IQSO che l'ICSO (Fig. 2.4).

**Fig. 2.4: ICSO vs IQSO, 2011**

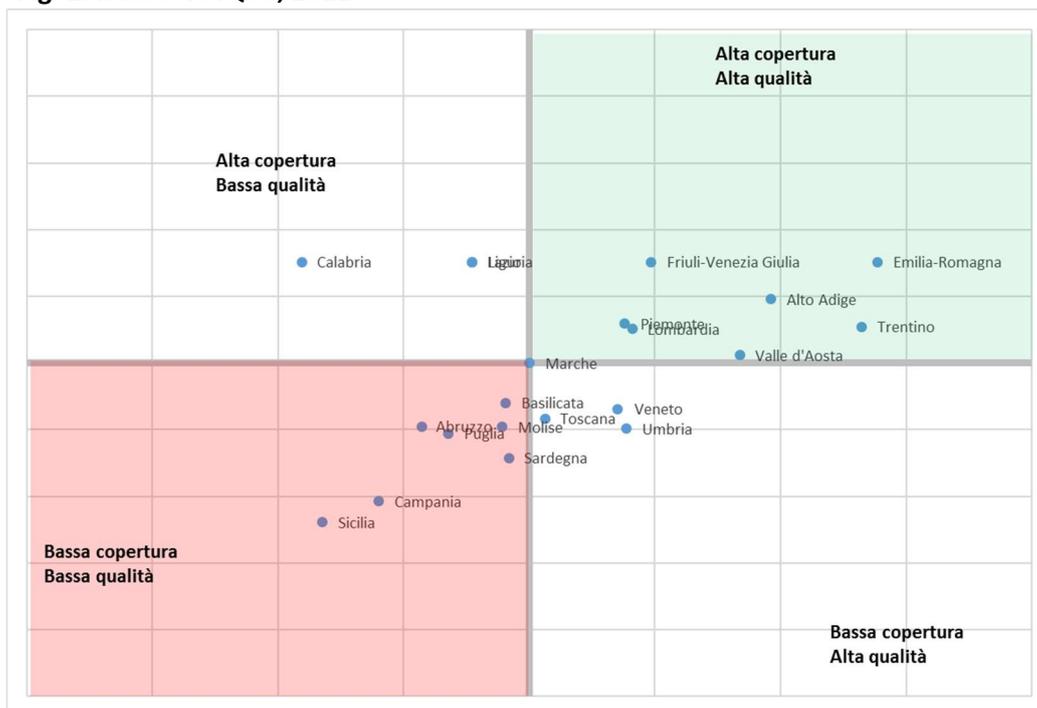

Nel 2012 il gruppo delle regioni con ICSO e IQSO elevato si sfoltisce, perdendo la Lombardia, che, tuttavia, mantiene un livello dei servizi ospedalieri qualitativamente elevato, e la Valle d'Aosta, che migra verso il quadrante con alto livello di copertura e basso livello di qualità (Fig. 2.5).

Si segnala come il Lazio, stabilmente nel quadrante in alto a sinistra nel 2010 e nel 2011, nel 2012 compie un passo consistente verso il quadrante delle regioni eccellenti, posizionandosi in corrispondenza dell'asse verticale (Fig. 2.5).

**Fig. 2.5: ICSO vs IQSO, 2012**

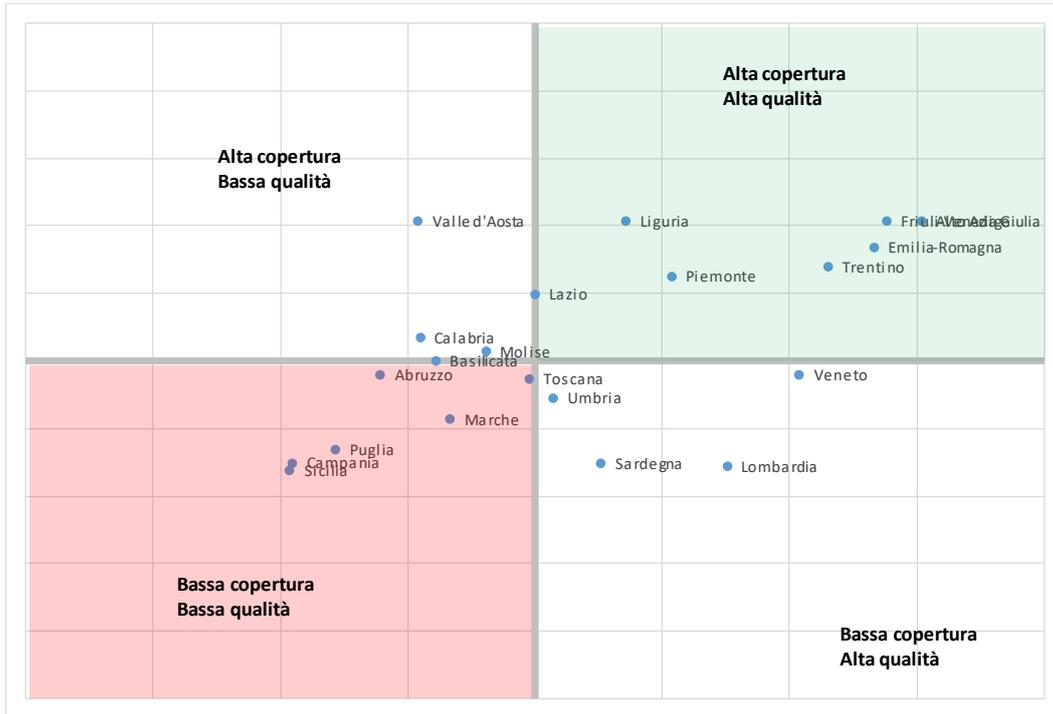

Infine la situazione 2013. Interessante notare come il Veneto entri nel quadrante delle regioni di eccellenza, la Basilicata esca da quello delle regioni con gli indici più bassi e la Toscana, dopo essersi affacciata al quadrante delle regioni con ICSO e IQSO più basso già nel 2012, in quest'anno si consolida all'interno di questo gruppo, unica regione non meridionale in quell'anno ad avere bassa copertura e bassa qualità (Fig. 2.6).

**Fig. 2.6: ICSO vs IQSO, 2013**

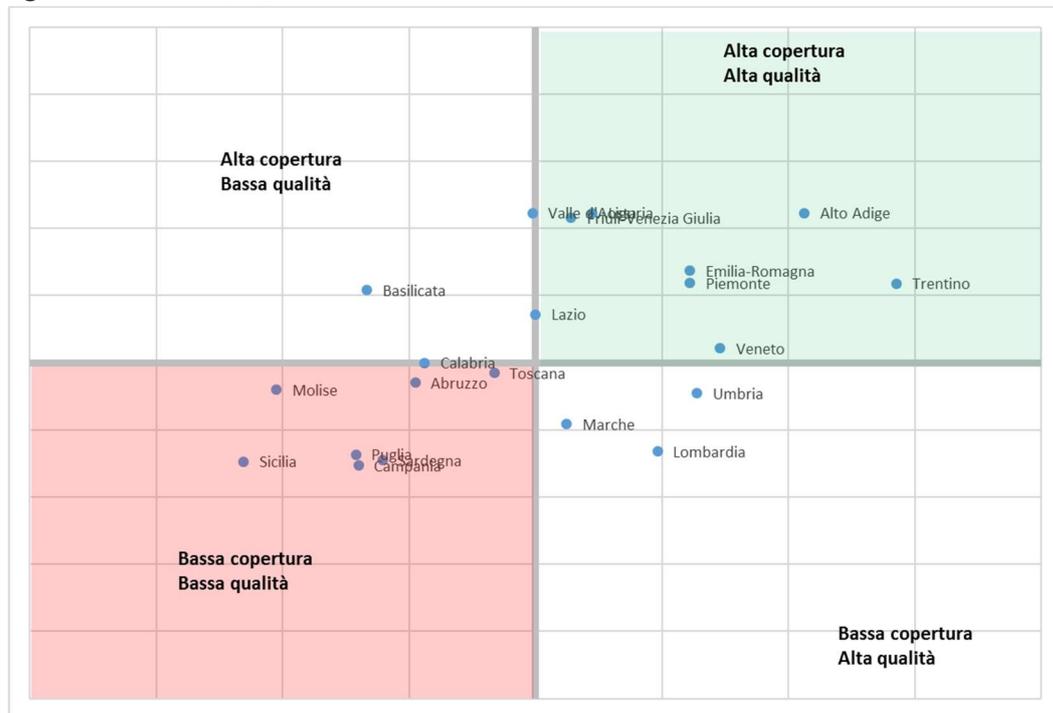

**2.4 Sintesi dei principali risultati**

Le principali osservazioni che è possibile ricavare incrociando i dati su copertura e qualità dei servizi ospedalieri sono:
- è forte, sotto entrambi i punti di vista, la differenza tra Nord e Sud Italia. Le regioni del Centro, invece, tendono a collocarsi a ridosso dei due gruppi;
- le regioni piccole (con popolazione inferiore ai 2 mln di abitanti) tendono a essere più numerose nel gruppo delle migliori, sebbene importante sia comunque la loro collocazione geografica;
- le regioni a statuto speciale sono quelle che maggiormente si concentrano, alternativamente, tra le migliori e le peggiori.

È interessante evidenziare come, salvo alcune eccezioni, le regioni che presentano le peggiori performance in termini sia di copertura del servizio, sia di qualità dello stesso, siano poi le stesse che, quando si guarda ai dati del saldo mobilità[9], presentano una situazione debitoria. Ciò, in parte, dipenderebbe proprio da quei pazienti che, per ricevere cure ospedaliere, emigrano verso altre regione (Tab. 2.3).

**Tab. 2.3: saldo mobilità anni 2008 - 2014**

| Regioni | riparto 2008 | riparto 2009 | riparto 2010 | riparto 2011 | riparto 2012 | riparto 2013 | riparto 2014 |
|---|---|---|---|---|---|---|---|
| PIEMONTE | -6.622 | 2.344 | 7.166 | 7.412 | 6.509 | -7.508 | -26.186 |
| VALLE D'AOSTA | -18.415 | -16.189 | -11.205 | -15.727 | -12.816 | -10.752 | -9.647 |
| LOMBARDIA | 420.547 | 461.011 | 464.268 | 433.028 | 457.499 | 555.183 | 533.960 |
| P. A. BOLZANO | 4.987 | 9.087 | 6.668 | 1.665 | 3.597 | 10.629 | 18.217 |
| P.A. TRENTO | -19.178 | -17.400 | -13.619 | -15.438 | -15.992 | -15.488 | -16.830 |
| VENETO | 104.612 | 79.705 | 84.635 | 93.741 | 95.180 | 75.790 | 75.357 |
| FRIULI VENEZIA GIULIA | 7.834 | 15.100 | 28.470 | 31.914 | 30.467 | 30.076 | 33.444 |
| LIGURIA | 1.618 | -28.797 | -37.866 | -40.029 | -43.967 | -56.743 | -51.770 |
| EMILIA ROMAGNA | 344.159 | 367.316 | 366.754 | 376.985 | 358.525 | 336.690 | 327.978 |
| TOSCANA | 107.011 | 100.731 | 99.435 | 121.082 | 131.927 | 132.294 | 151.214 |
| UMBRIA | 777 | -2.946 | 9.722 | 9.886 | 2.649 | 9.411 | 3.295 |
| MARCHE | -44.491 | -46.099 | -32.507 | -19.644 | -22.472 | -33.677 | -46.146 |
| LAZIO | -62.869 | -108.361 | -99.562 | -55.149 | -118.979 | -199.100 | -201.575 |
| ABRUZZO | -6.334 | -34.647 | -71.692 | -124.377 | -101.407 | -69.559 | -70.715 |
| MOLISE | 36.328 | 36.353 | 28.196 | 37.548 | 35.839 | 30.109 | 25.722 |
| CAMPANIA | -313.066 | -318.434 | -308.213 | -331.542 | -299.132 | -310.810 | -270.403 |
| PUGLIA | -222.316 | -103.002 | -139.981 | -177.009 | -171.737 | -180.058 | -187.265 |
| BASILICATA | -30.188 | -37.429 | -40.348 | -33.211 | -19.140 | -19.111 | -38.796 |
| CALABRIA | -224.172 | -239.292 | -257.493 | -233.992 | -250.009 | -251.654 | -251.687 |
| SICILIA | -209.720 | -224.162 | -203.788 | -208.590 | -203.844 | -188.774 | -161.682 |
| SARDEGNA | -70.476 | -74.401 | -67.050 | -50.956 | -64.796 | -68.787 | -70.564 |
| B. GESU' | 178.974 | 148.781 | 149.903 | 157.854 | 165.859 | 192.720 | 194.527 |
| ACISMOM | 21.001 | 30.730 | 38.108 | 34.549 | 36.241 | 39.120 | 39.552 |
| ITALIA | 0 | 0 | 0 | 0 | 0 | 0 | 0 |

*Valori migliaia di euro*

Fonte: elaborazione Agenas su dati intese Stato-Regioni-riparto disponibilità finanziarie anni 2008-2014

---

[9] Si tratta delle somme che le regioni si versano reciprocamente in relazione al numero di soggetti che immigrano o emigrano da queste per ricevere cure ospedaliere ma anche servizi di tipo distrettuale, si pensi, per esempio, al caso degli studenti o dei lavoratori fuori sede.

## 3. La valutazione dei servizi di assistenza distrettuale regionali: 2010 – 2013

La spesa per servizi assistenziali distrettuali rappresenta all'incirca il 50% della spesa sanitaria complessiva (fatte le dovute distinzioni tra regioni). Assieme alla spesa per servizi ospedalieri, è la principale componente della spesa sanitaria regionale.

Partendo dalle informazioni disponibili nella banca dati HFA, è qui proposta una valutazione di questa tipologia di servizi nel periodo 2010 – 2013. L'analisi si compone, da un lato della valutazione della copertura dei servizi stessi rispetto alla popolazione di riferimento, dall'altro della valutazione della loro qualità.

Con riferimento alla costruzione dell'Indice di Copertura del Servizio Distrettuale (ICSD), le variabili prese in considerazione sono:

- tasso servizi di guardia medica (per mille abitanti);
- tasso medici generici (per 10 mila abitanti);
- tasso pediatri di base (per 10 mila abitanti);
- tasso posti letto residenziali per funzione socio-sanitaria (per 10 mila abitanti);
- tasso strutture sanitarie distrettuali (per 10 mila abitanti).

Per quel che concerne, invece, la costruzione dell'Indice di Qualità del Servizio Distrettuale (IQSD), le variabili prese in considerazione sono:

- numero medio di medici di guardia medica per servizio;
- medici generici con più di 1.500 assistiti (in percentuale dei medici generici);
- pediatri con più di 800 assistiti (in percentuale dei pediatri).

Il collegamento di queste tre variabili al livello di qualità del servizio offerto è così spiegato:

- maggiore è il numero di medici per servizio di guardia e minore è il numero medio di ore di servizio (stando ai dati HFA 2010-2013, la correlazione tra queste due variabili è pari a -0,75). Da ciò si può supporre che il servizio sia migliore tanto più numerosi sono i medici di guardia per turno, dal momento che i loro turni sono più brevi;
- quando il numero di medici di base e pediatri che assistono molti pazienti è elevato, si può supporre che la qualità del servizio reso sia mediamente più bassa, anche semplicemente per il fatto che i tempi di attesa necessari per avere un appuntamento si dilatano.

Partendo da questi tassi, la sintesi, con costruzione di un unico ICSD ed un unico IQSD, è realizzata ricorrendo al modello MNCP, già utilizzato in precedenza per valutare i servizi ospedalieri.

### 3.1 L'Indice di Copertura del Servizio Distrettuale

L'ICSD si configura come un indice di sintesi riferito alla capacità regionale di erogare determinati servizi distrettuali (vedi lista paragrafo precedente) in rapporto alla popolazione locale. Il valore dell'indice è compreso tra 0 (valore minimo) e 1 (copertura massima).

Guardando al valore medio annuo degli ICSD regionali, si evidenzia come, tra il 2010 ed il 2013, questo si incrementa costantemente, passando da 0,66 a 0,73 (Fig. 3.1).

Si sottolinea che, nel tempo, è andato leggermente aumentando il valore della deviazione standard (da poco meno di 0,27 a 0,3), indice del fatto che la eterogeneità dei punteggi ottenuti dalle singole regioni è leggermente aumentata.

**Fig. 3.1: ICSD nel periodo 2010 – 2013: media valori regionali per anno**

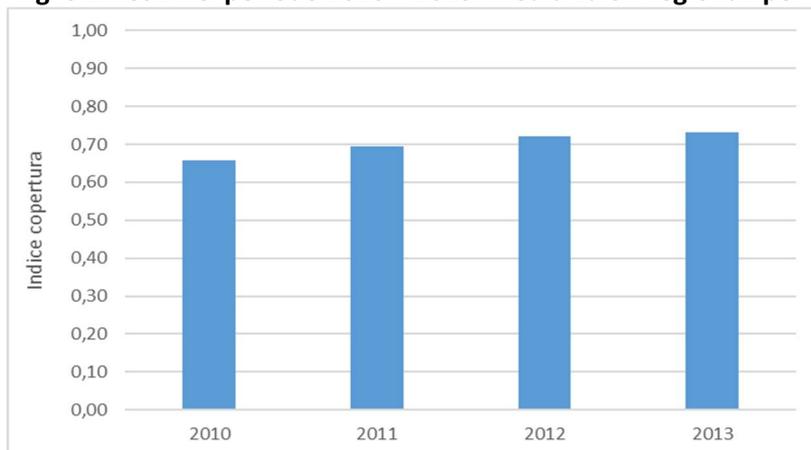

Andando a guardare la situazione dettagliata per singola regione, emergono interessanti differenze. Le regioni che, abbastanza stabilmente, mantengono un ICSD elevato durante tutto il periodo considerato sono: l'Emilia Romagna, la Sardegna, la Liguria, la Sicilia, la Toscana e la Valle d'Aosta (Tab. 3.1).

Tra le regioni che, al contrario, restano agli ultimissimi posti durante tutto il periodo ci sono Alto Adige e Campania (Tab. 3.1).

Le regioni che, nel tempo, migliorano la propria performance sono Molise, Umbria e Basilicata (Tab. 3.1).

Il Lazio è la regione che, invece, peggiora in maniera evidente, in particolare tra 2011 e anni successivi (Tab. 3.1).

A livello di macro aree regionali, non esiste una chiara "superiorità" o "inferiorità" dell'una rispetto all'altra, come, invece, visto per i servizi ospedalieri (Tab. 3.1).

**Tab. 3.1: ICSD regionali nel periodo 2010 - 2013**

| Regione | 2010 | 2011 | 2012 | 2013 |
|---|---|---|---|---|
| Abruzzo | 0,735 | 1,000 | 1,000 | 0,819 |
| Alto Adige | 0,188 | 0,228 | 0,226 | 0,224 |
| Basilicata | 0,571 | 0,820 | 1,000 | 1,000 |
| Calabria | 0,683 | 0,795 | 0,792 | 1,000 |
| Campania | 0,101 | 0,054 | 0,065 | 0,058 |
| Emilia-Romagna | 0,931 | 0,973 | 1,000 | 0,847 |
| Friuli-Venezia Giulia | 0,649 | 0,716 | 0,716 | 0,618 |
| Lazio | 0,899 | 1,000 | 0,153 | 0,108 |
| Liguria | 1,000 | 0,779 | 1,000 | 0,875 |
| Lombardia | 0,459 | 0,453 | 0,512 | 0,514 |
| Marche | 0,355 | 0,309 | 0,369 | 0,497 |
| Molise | 0,630 | 1,000 | 1,000 | 1,000 |
| Piemonte | 0,471 | 0,549 | 0,762 | 0,719 |
| Puglia | 0,573 | 0,603 | 0,508 | 0,771 |
| Sardegna | 1,000 | 0,940 | 1,000 | 0,992 |
| Sicilia | 1,000 | 0,785 | 1,000 | 0,968 |
| Toscana | 0,967 | 0,961 | 0,926 | 0,940 |
| Trentino | 0,631 | 0,753 | 0,927 | 0,907 |
| Umbria | 0,375 | 0,322 | 0,581 | 1,000 |
| Valle d'Aosta | 1,000 | 0,977 | 1,000 | 0,941 |
| Veneto | 0,606 | 0,571 | 0,627 | 0,594 |

**Legenda:** la scala cromatica utilizzata va dal rosso intenso (ICSD basso) al verde intenso (ICSD alto).

## 3.2 L'indice di Qualità del Servizio Distrettuale

Avendo visto le differenze cronologiche e territoriali riguardanti la quantità di servizi di assistenza distrettuale offerti rispetto alla popolazione, utile è guardare anche a quello che succede da un punto di vista più prettamente qualitativo. L'IQSD sintetizza in un unico indice la qualità determinata sulla base di alcuni elementi riferiti, in particolare, ai servizi di medici di base, pediatri e guardie mediche. Il suo valore è compreso tra 0 (qualità scarsa) a 1 (qualità massima).

Il livello medio regionale dell'IQSD cala tendenzialmente nel tempo, passando da 0,71 a 0,64, ad indicare che, se è vero che da un lato è aumentata la copertura (Fig. 3.1), dall'altro ciò ha finito col determinare un abbassamento della qualità del servizio offerto (Fig. 3.2).

Deviazione standard leggermente ridotta nel tempo (da 0,22 a 0,19), quindi il peggioramento della qualità 2013 vede i valori regionali molto vicini a quello medio, più di quanto non accada ad inizio periodo.

**Fig. 3.2: IQSD nel periodo 2010 – 2013: media valori regionali per anno**

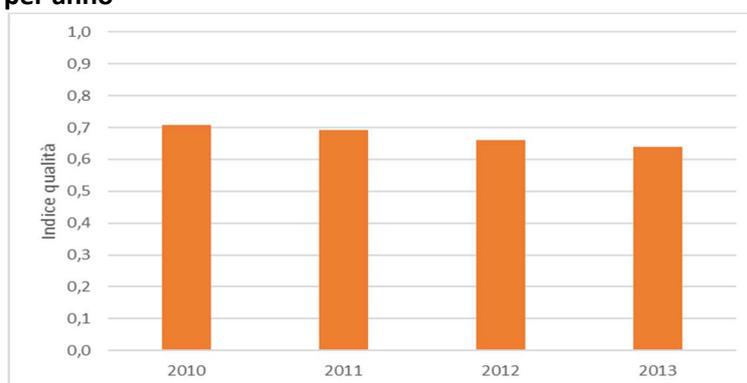

Per quanto concerne il dettaglio regionale, si evidenzia come le regioni che durante tutto l'arco di tempo considerato registrano il livello qualitativo più elevato dei servizi di assistenza distrettuale sono Abruzzo, Campania, Sardegna, Umbria e, un po' meno, Emilia Romagna e Sicilia (Tab. 3.2).

Tra le regioni che presentano l'IQSD più basso ci sono Friuli Venezia Giulia, Trentino e Valle d'Aosta (Tab. 3.2). Lazio e Molise peggiorano la performance nel tempo, mentre la Liguria la migliora (Tab. 3.2).

**Tab. 3.2: IQSD regionali nel periodo 2010 – 2013**

| Regione | 2010 | 2011 | 2012 | 2013 |
|---|---|---|---|---|
| Abruzzo | 0,922 | 0,914 | 0,898 | 0,907 |
| Alto Adige | 0,473 | 0,452 | 0,465 | 0,481 |
| Basilicata | 1,000 | 0,898 | 0,899 | 0,617 |
| Calabria | 0,629 | 0,636 | 0,598 | 0,634 |
| Campania | 0,947 | 1,000 | 0,897 | 0,942 |
| Emilia-Romagna | 0,745 | 0,804 | 0,787 | 0,779 |
| Friuli-Venezia Giulia | 0,305 | 0,409 | 0,363 | 0,328 |
| Lazio | 0,707 | 0,836 | 0,588 | 0,561 |
| Liguria | 0,659 | 0,621 | 0,783 | 0,754 |
| Lombardia | 0,635 | 0,641 | 0,616 | 0,599 |
| Marche | 0,646 | 0,531 | 0,464 | 0,509 |
| Molise | 0,817 | 0,520 | 0,475 | 0,526 |
| Piemonte | 0,540 | 0,429 | 0,469 | 0,435 |
| Puglia | 0,870 | 0,872 | 0,878 | 0,828 |
| Sardegna | 1,000 | 0,975 | 0,964 | 0,866 |
| Sicilia | 0,982 | 0,972 | 0,859 | 0,790 |
| Toscana | 0,667 | 0,610 | 0,580 | 0,587 |
| Trentino | 0,408 | 0,270 | 0,236 | 0,288 |
| Umbria | 1,000 | 0,994 | 0,971 | 0,904 |
| Valle d'Aosta | 0,234 | 0,270 | 0,338 | 0,442 |
| Veneto | 0,706 | 0,876 | 0,744 | 0,672 |

**Legenda:** la scala cromatica utilizzata va dal rosso intenso (ICSD basso) al verde intenso (ICSD alto).

**3.3 Copertura e Qualità**

Dopo aver guardato i due indicatori separatamente, si procede incrociando le informazioni, in modo tale da valutare se e per quali regioni un ICSD elevato si accompagna ad un IQSD soddisfacente e viceversa. Per far questo ci si avvale di analisi grafica. Grazie ad essa è possibile suddividere il piano cartesiano in quattro quadranti e le regioni in altrettanti gruppi[10]:

- quadrante in alto a destra – "Alta copertura, Alta qualità": è l'area in cui si trovano le regioni che ottengono le migliori performance sia dal punto di vista della capacità di copertura del servizio di assistenza distrettuale, sia in termini di qualità del servizio offerto;
- quadrante in basso a destra – "Bassa copertura, Alta qualità": è l'area in cui si trovano le regioni che, pur ottenendo buone performance dal punto di vista della qualità del servizio offerto, non sono in grado di garantire livelli di copertura paragonabili a quelli delle regioni presenti nel quadrante superiore;
- quadrante in basso a sinistra – "Bassa copertura, Bassa qualità": è l'area in cui si trovano le regioni che fanno peggio in termini di servizi offerti, avendo un basso ICSD ed un altrettanto basso IQSD;
- quadrante in alto a sinistra – "Alta copertura, Bassa qualità": è l'area in cui si trovano le regioni che, seppur in grado di offrire una buona copertura di servizi distrettuali rispetto alla popolazione, non eccellono per quel che riguarda la qualità di quanto offerto.

Tra le regioni di eccellenza nel 2010 ci sono le isole maggiori, l'Emilia Romagna e l'Abruzzo (Fig. 3.3).

Tra le regioni con le performance peggiori ci sono le Marche, la Lombardia, il Piemonte e l'Alto Adige (Fig. 3.3).

Molte regioni si posizionano a ridosso degli assi, a significare che la maggior parte di esse presenta valori molto vicini e, dunque, il fatto di ricadere in uno o nell'altro quadrante è dovuto a differenze minime degli indici (Fig. 3.3).

Interessante evidenziare come, confrontando questa analisi grafica con quella fatta per i servizi ospedalieri, la disposizione che assumono i punti è più casuale, con un numero più uniforme di unità all'interno del singolo quadrante. In questo caso, quindi, è molto meno evidente la relazione tra quantità e qualità del servizio offerto.

---

[10] Gli assi delle ascisse e delle ordinate vengono fatti passare in corrispondenza dei valori mediani di ICSD e IQSD, in modo tale da "spaccare" i campioni in due, incrociando poi i risultati.

**Fig. 3.3: ICSD vs IQSD, 2010**

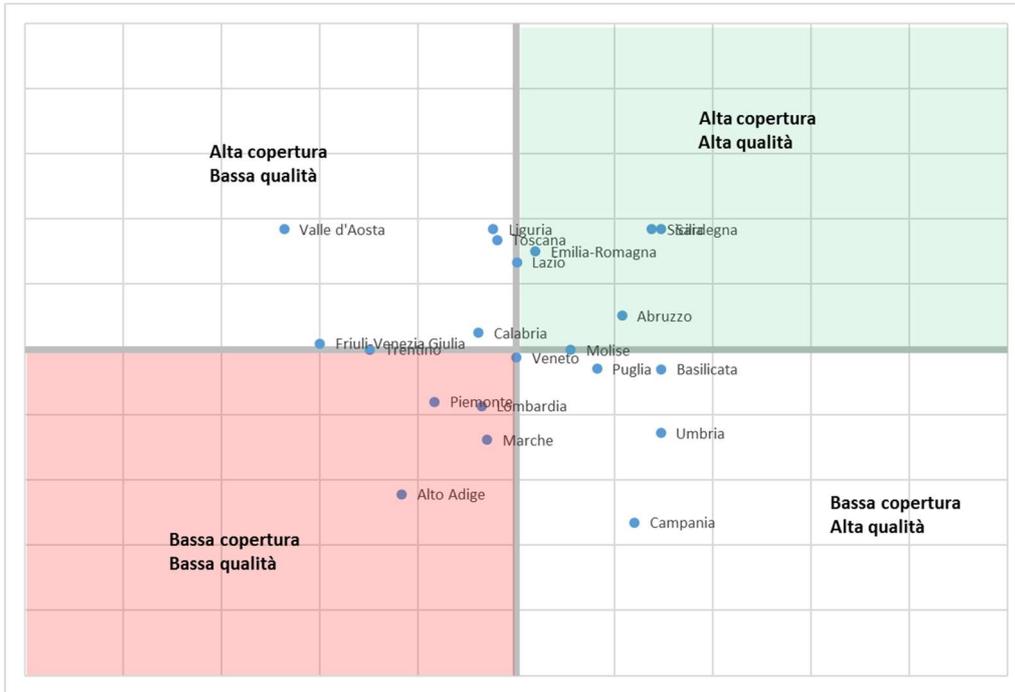

La situazione del 2011 è simile a quella dell'anno precedente. Tra le principali differenze, il fatto che si "schiacciano" verso l'origine degli assi le regioni del quadrante in alto a destra, a segnalare una riduzione delle distanze rispetto alle altre regioni, più per qual che concerne la copertura del servizio, che per la qualità. La Campania, già distante dalle altre regioni per quel che riguarda la copertura, si distanzia ancor di più lungo quest'asse (Fig. 3.4).

**Fig. 3.4: ICSD vs IQSD, 2011**

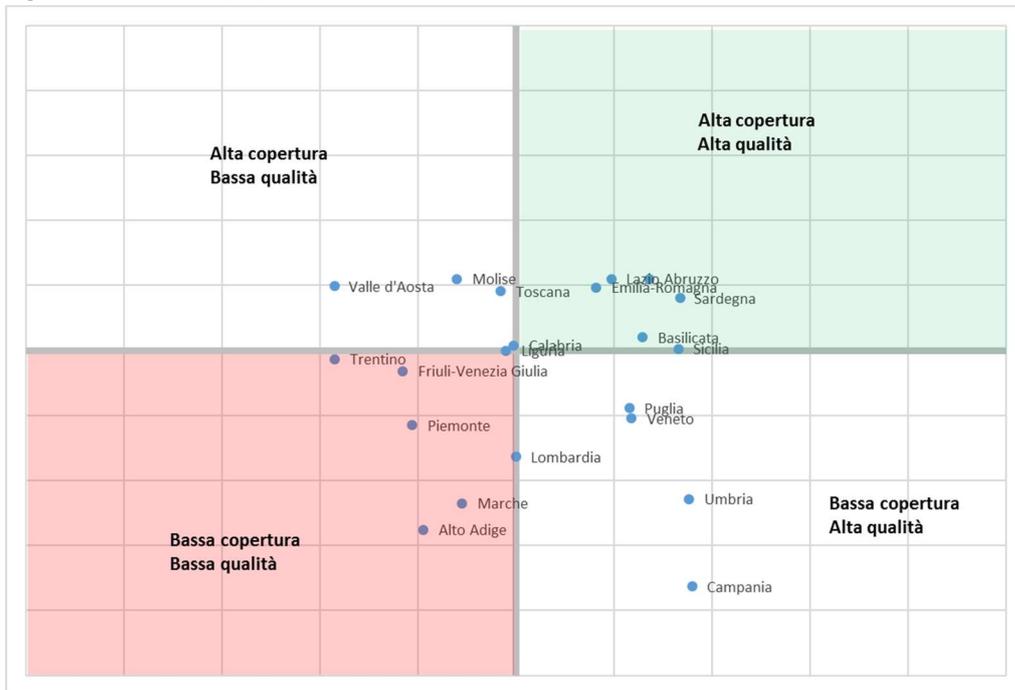

Nel 2012, il gruppo delle regioni con ICSD e IQSD elevati (le due isole, Liguria, Emilia Romagna e Basilicata) si concentra attorno a stessi valori. (Fig. 3.5).

Si segnala come il Lazio, stabilmente nel quadrante in alto a destra nel 2010 e nel 2011, nel 2012 compie un passo consistente verso il quadrante delle regioni con le performance peggiori, soprattutto a causa di una forte perdita per quel che concerne la copertura (Fig. 3.5).

**Fig. 3.5: ICSD vs IQSD, 2012**

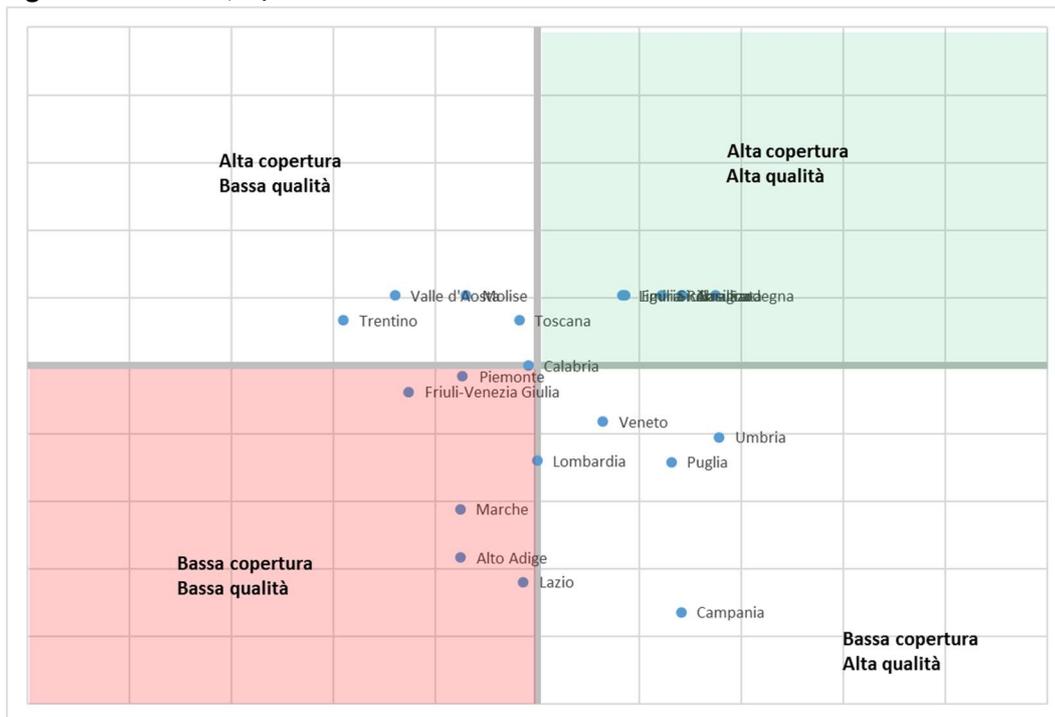

Infine la situazione 2013. Interessante notare come entrino (o rientrino) nel gruppo delle migliori l'Umbria e la Calabria, con l'Emilia Romagna che peggiora un po' la performance rispetto agli anni precedenti. In generale, si assiste ad uno "schiacciamento" verso l'asse delle x, quello orizzontale, il che significa che le differenze tra regioni sono legate per lo più alla qualità del servizio offerto e molto meno alla sua copertura rispetto alla popolazione (fatta eccezione per, nell'ordine, Alto Adige, Lazio e, soprattutto, Campania) (Fig. 3.6).

**Fig. 3.6: ICSD vs IQSD, 2013**

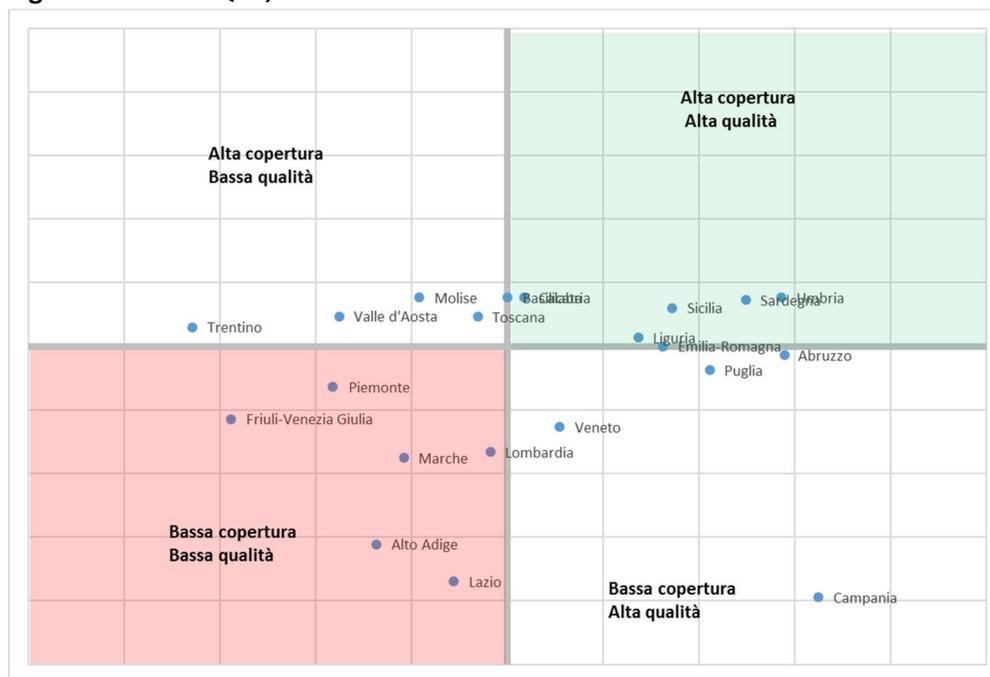

## 3.4 Sintesi dei principali risultati

Nella valutazione dei risultati ottenuti, interessante è guardare quello che accade incrociando i dati sulla copertura dei servizi ospedalieri con quelli relativi alla copertura dei servizi sanitari distrettuali. A questo proposito, è possibile distinguere 4 diversi gruppi di regioni:
- "Alta copertura dei servizi ospedalieri e di quelli di assistenza distrettuale": Emilia Romagna, Liguria e Valle d'Aosta sono le regioni che riescono a garantire la miglior copertura relative ai principali servizi sanitari;
- "Bassa copertura servizi ospedalieri, alta copertura dei servizi di assistenza distrettuale": le isole maggiori, l'Abruzzo e la Toscana sono le regioni che, pur non avendo ottima copertura dei servizi ospedalieri, compensano con quelli territoriali;
- "Alta copertura servizi ospedalieri, bassa copertura dei servizi di assistenza distrettuale": Piemonte, Lazio, Friuli Venezia Giulia e Alto Adige sono le regioni che, pur eccellendo sul piano dei servizi ospedalieri offerti, accusano ritardo per quel che riguarda i servizi di assistenza distrettuale;
- "Bassa copertura dei servizi ospedalieri e di quelli di assistenza territoriale e distrettuale": Veneto, Lombardia, Puglia, Marche, Umbria e, molto distaccata, la Campania sono le regioni che hanno problemi nel riuscire ad erogare un livello sufficiente, in termini quantitativi, dei propri servizi sanitari alla popolazione.

Intermedia la situazione delle regioni che si piazzano lungo gli assi (Fig. 3.7).

**Fig. 3.7: copertura dei servizi ospedalieri vs copertura dei servizi distrettuali: valori medi regionali, periodo 2010-2013**

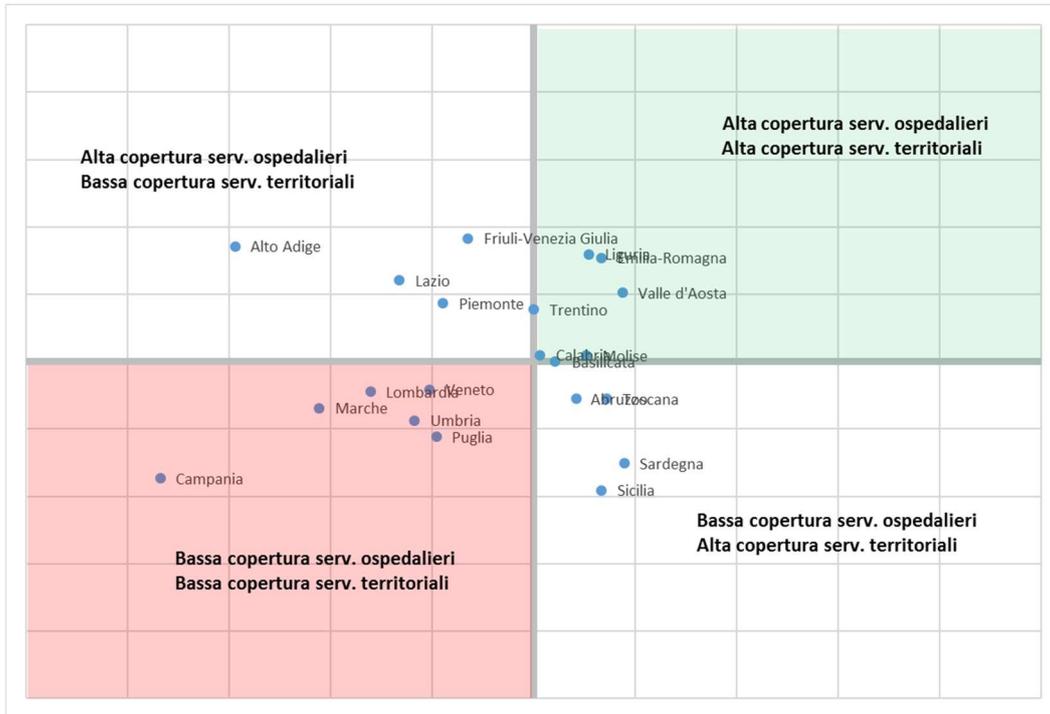

Quando, invece, si va a guardare alla qualità dei servizi offerti, i quattro gruppi di regioni che si ottengono sono:
- "Alta qualità dei servizi ospedalieri e di quelli di assistenza distrettuale": Emilia Romagna, Liguria, Veneto e Umbria sono le regioni che riescono a garantire la miglior qualità relativamente ai principali servizi sanitari;
- "Bassa qualità dei servizi territoriali, alta qualità dei servizi di assistenza distrettuale": le isole maggiori, l'Abruzzo, la Puglia, la Basilicata e la Campania sono le regioni che, pur non avendo ottima qualità dei servizi ospedalieri, compensano con quelli territoriali;
- "Alta qualità dei servizi ospedalieri, bassa qualità dei servizi di assistenza distrettuale": Piemonte, Alto Adige, Trentino, Friuli Venezia Giulia, Lombardia e Valle d'Aosta sono le regioni che, pur eccellendo sul piano della qualità dei servizi ospedalieri offerti, accusano un ritardo per quel che riguarda i servizi di assistenza distrettuale;
- "Bassa qualità dei servizi ospedalieri e di quelli di assistenza distrettuale": Marche, Molise e Calabria sono le regioni che hanno problemi nel riuscire ad erogare un livello qualitativamente sufficiente di servizi sanitari.

Intermedia la situazione delle regioni che si piazzano lungo gli assi (Fig. 3.8).

**Fig. 3.8: qualità dei servizi ospedalieri vs qualità dei servizi distrettuali: valori medi regionali, periodo 2010-2013**

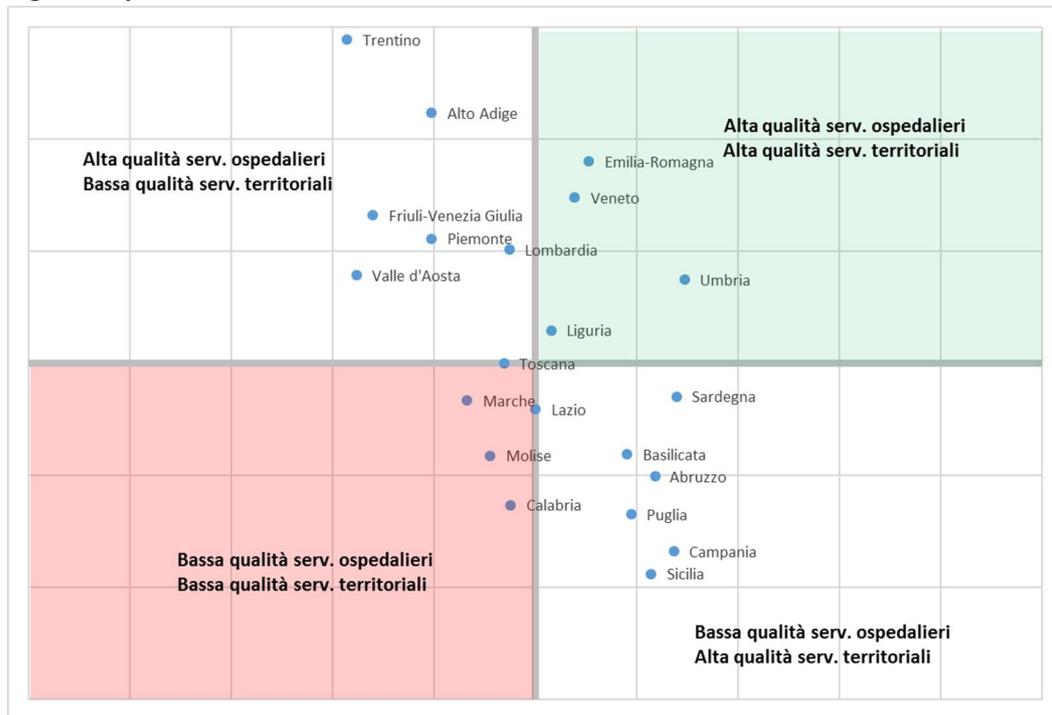

**4. L'efficienza dei servizi sanitari regionali**

Fino a questo momento, parlando di copertura e qualità dei servizi offerti, si è fatto riferimento all'efficacia della sanità regionale, intesa come capacità, da parte delle regioni, di soddisfare la domanda proveniente dalla popolazione. Tale misurazione è fatta confrontando i livelli di alcuni servizi base erogati. Sebbene, spesso, si tenda a sovrapporre i concetti, l'efficacia è altra cosa rispetto all'efficienza. Quest'ultima consiste nella capacità di offrire un certo bene o servizio (a prescindere dal fatto che esso sia ritenuto soddisfacente, insufficiente o eccessivo) con l'impiego più basso possibile di risorse. È proprio la valutazione dell'efficienza dei servizi sanitari l'oggetto delle prossime pagine.

Nel momento in cui si parla di efficienza, bisogna innanzitutto considerare i costi del servizio. A differenza di quanto fatto in apertura di questo report, è qui esaminato l'andamento del costo pro capite regionale nel periodo in esame, in modo da pesare la spesa totale in base alla popolazione. Questo andamento è caratterizzato da una sostanziale stabilità tra 2010 e 2011, con successiva riduzione: la spesa pro capite sanitaria media regionale, che nel 2011 raggiunge i 1.923 €, scende, nel 2013, a 1.885 €. Quasi 50 € in meno in 3 anni (Fig. 4.1).

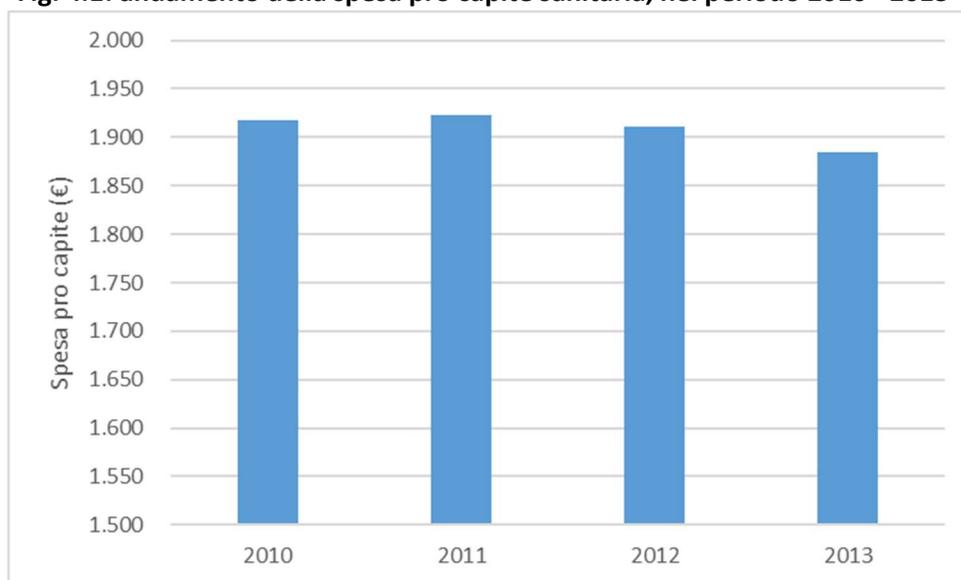

**Fig. 4.1: andamento della spesa pro capite sanitaria, nel periodo 2010 - 2013**

La situazione, tuttavia, non è omogenea su tutto il territorio. In Molise, per esempio, la spesa pro capite 2013 è più alta di quella che si registra negli anni precedenti, così come in regioni quali la Sardegna o l'Alto Adige, pur essendo la spesa pro capite 2013 più bassa di quella 2012, resta comunque più elevata rispetto a quella degli anni 2010 e 2011 (Tab. 4.1).

A prescindere dall'andamento nel tempo, è interessante notare come in regioni quali Valle d'Aosta, Liguria, Alto Adige, Friuli Venezia Giulia e Molise la spesa pro capite sanitaria è costantemente al di sopra dei 2 mila € annui lungo tutto il periodo considerato, mentre in regioni come Calabria e Campania, nel 2013 in particolare, questo valore scende al di sotto dei 1.700 (Tab. 4.1). Tra le ragioni che spiegano queste differenze, lo status di regione a statuto speciale e la presenza, più numerosa che altrove, di soggetti anziani che, in quanto tali, richiedono un maggior numero di cure (vedi Liguria in particolare).

**Tab. 4.1: andamento della spesa pro capite sanitaria, nel periodo 2010 – 2013, per regione**

| Regione | 2010 | 2011 | 2012 | 2013 |
|---|---|---|---|---|
| Abruzzo | 1.777 | 1.798 | 1.776 | 1.739 |
| Alto Adige | 2.221 | 2.235 | 2.281 | 2.241 |
| Basilicata | 1.865 | 1.888 | 1.810 | 1.817 |
| Calabria | 1.778 | 1.755 | 1.724 | 1.696 |
| Campania | 1.790 | 1.752 | 1.707 | 1.667 |
| Emilia-Romagna | 1.854 | 1.870 | 1.890 | 1.857 |
| Friuli-Venezia Giulia | 2.007 | 2.066 | 2.056 | 2.020 |
| Lazio | 1.995 | 1.994 | 2.052 | 1.954 |
| Liguria | 2.072 | 2.066 | 2.039 | 2.006 |
| Lombardia | 1.826 | 1.853 | 1.823 | 1.819 |
| Marche | 1.839 | 1.827 | 1.786 | 1.763 |
| Molise | 2.159 | 2.092 | 2.094 | 2.220 |
| Piemonte | 1.923 | 1.908 | 1.876 | 1.831 |
| Puglia | 1.843 | 1.797 | 1.763 | 1.766 |
| Sardegna | 1.958 | 1.990 | 2.060 | 2.020 |
| Sicilia | 1.776 | 1.775 | 1.737 | 1.718 |
| Toscana | 1.891 | 1.909 | 1.868 | 1.812 |
| Trentino | 1.897 | 1.943 | 1.967 | 1.931 |
| Umbria | 1.840 | 1.861 | 1.863 | 1.838 |
| Valle d'Aosta | 2.194 | 2.244 | 2.216 | 2.153 |
| Veneto | 1.754 | 1.750 | 1.733 | 1.709 |

**Legenda:** la scala cromatica utilizzata va dal rosso intenso (spesa pro capite bassa) al verde intenso (spesa pro capite alta).

Un'informazione completa rispetto ai costi della sanità dovrebbe tener conto non soltanto del valore pro capite, ma anche della percentuale di spesa che è direttamente a carico delle famiglie, attraverso il pagamento di ticket. In tal senso si nota, per esempio, che in Veneto, una delle regioni con il valore più basso di spesa pro capite complessiva, è molto elevata la quota di spesa a carico dei privati (circa il 27/28% del totale), mentre in Molise, nonostante la spesa pro capite elevata, resta bassa la quota a carico delle famiglie (18/19%) (Tab. 4.2).

La regione in cui la quota di partecipazione delle famiglie è più bassa è la Campania, che è anche tra le regioni che registrano anche il livello più basso di spesa pro capite. La regione in cui la percentuale di spesa a carico delle famiglie è più alta è, invece, il Friuli Venezia Giulia, regione che, al contrario della Campania, è tra quelle con un livello di spesa pro capite più elevato (Tab. 4.2).

**Tab. 4.2: % spesa sanitaria a carico delle famiglie, nel periodo 2010 – 2013, per regione**

| Regione | 2010 | 2011 | 2012 | 2013 |
|---|---|---|---|---|
| Abruzzo | 21,05 | 21,80 | 22,11 | 21,70 |
| Alto Adige | 22,14 | 23,60 | 22,04 | 21,25 |
| Basilicata | 19,52 | 20,67 | 22,51 | 22,82 |
| Calabria | 19,76 | 21,08 | 22,01 | 21,54 |
| Campania | 14,97 | 16,30 | 17,15 | 17,84 |
| Emilia-Romagna | 25,89 | 27,26 | 27,37 | 27,59 |
| Friuli-Venezia Giulia | 30,42 | 31,57 | 29,50 | 28,76 |
| Lazio | 21,52 | 22,91 | 21,46 | 20,77 |
| Liguria | 19,58 | 20,56 | 21,55 | 21,99 |
| Lombardia | 23,27 | 24,67 | 24,94 | 25,45 |
| Marche | 22,82 | 24,21 | 23,63 | 23,61 |
| Molise | 18,10 | 19,04 | 17,96 | 17,65 |
| Piemonte | 21,85 | 23,14 | 24,31 | 25,37 |
| Puglia | 18,29 | 20,09 | 19,88 | 20,08 |
| Sardegna | 18,11 | 19,12 | 19,30 | 18,63 |
| Sicilia | 17,48 | 18,20 | 18,99 | 18,37 |
| Toscana | 21,42 | 22,93 | 23,14 | 23,79 |
| Trentino | 23,02 | 24,12 | 23,89 | 23,98 |
| Umbria | 22,85 | 23,41 | 22,21 | 21,81 |
| Valle d'Aosta | 29,29 | 30,48 | 29,20 | 29,63 |
| Veneto | 26,57 | 28,45 | 27,66 | 26,86 |

**Legenda:** la scala cromatica utilizzata va dal rosso intenso (% bassa) al verde intenso (% alta).

**4.1 La valutazione dell'efficienza**

Partendo dai risultati delle valutazioni riferite a copertura e qualità dei servizi ospedalieri (ICSO e IQSO) e di quelli distrettuali (ICSD e IQSD) e conoscendo i costi pro capite regionali che garantiscono quei servizi, è possibile realizzare due valutazioni di efficienza, ognuna delle quali riferita a ciascuna delle due tipologie di assistenza. Per fare questo, è necessario "spaccare" i costi pro capite complessivi in due componenti, ognuna delle quali riferita ai due servizi analizzati. Per far questo, si ricorre alle informazioni, su base annua, del NINS (Fig. 4.2)[11].

**Fig. 4.2: ripartizione % della spesa sanitaria nazionale tra livelli macro assistenziali, periodo 2008 - 2013**

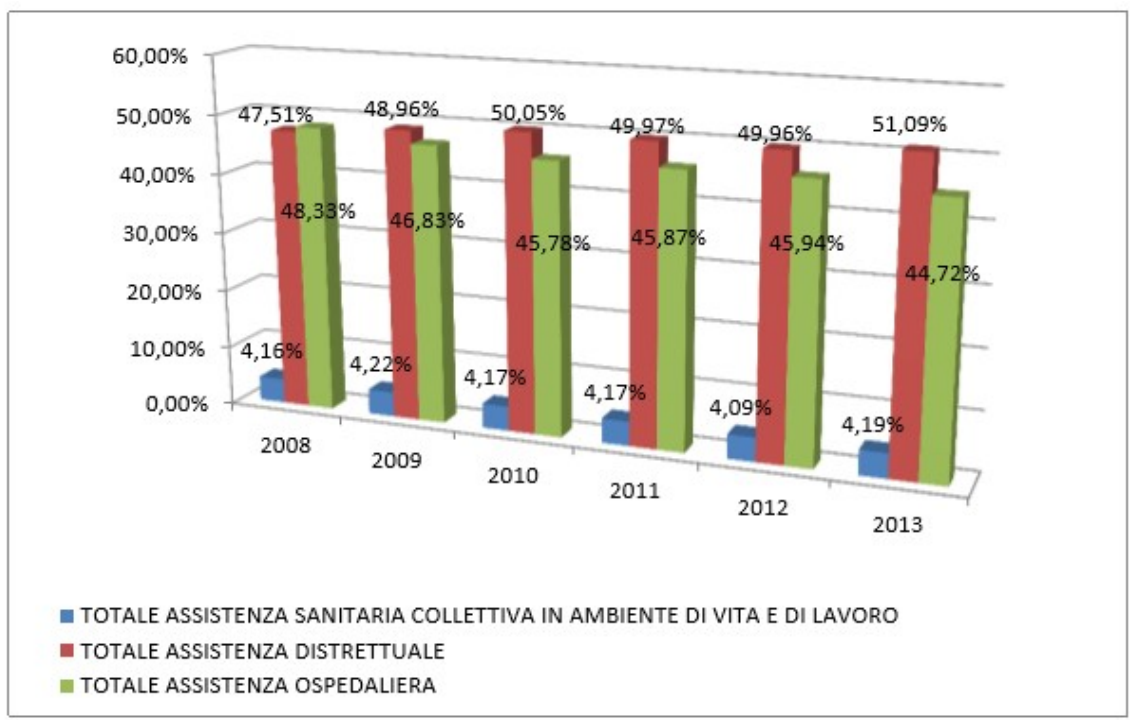

Fonte: elaborazione Agenas su modelli LA 2008-2013 (NSIS)

Per quanto concerne il modello utilizzato per la valutazione dell'efficienza, si utilizza un modello Data Envelopment Analysis (DEA) Input Oriented con rendimenti di scala variabili (Banker et al., 1984). Le componenti di costo pro capite riferite all'assistenza ospedaliera e a quella distrettuale sono valutate in base ai rispettivi livelli di copertura e di qualità del servizio. Gli indici così ottenuti oscillano tra 1 (efficienza massima) e 0 (efficienza minima).

Per quanto concerne l'efficienza nell'erogazione dei servizi ospedalieri, la regione che ha le migliori performance lungo tutto l'arco di tempo considerato è l'Emilia Romagna. Seguono, a breve distanza, il Trentino e il Friuli Venezia Giulia. Sono queste, dunque, le regioni che, stante il livello di servizi ospedalieri offerti, riescono a farlo ai costi più bassi (Tab. 4.3).

Pessime le performance di Sicilia e Campania (Tab. 4.3).

Da segnalare come molte regioni del Nord Italia, con la Basilicata, migliorino la loro efficienza nel tempo. Lazio e Lombardia, invece, affrontano un peggioramento, a partire dal 2012 (Tab. 4.3).

---

[11] Data la sostanziale irrilevanza dell'assistenza sanitaria collettiva in ambiente di vita e lavoro (Fig. 3.2), la quota di spesa ad essa riferita è ridistribuita proporzionalmente sulle altre due tipologie di assistenza.

**Tab. 4.3: l'efficienza nell'erogazione dei servizi ospedalieri, periodo 2010 – 2013, per regione**

| Regione | 2010 | 2011 | 2012 | 2013 |
|---|---|---|---|---|
| Abruzzo | 0,55 | 0,52 | 0,56 | 0,55 |
| Basilicata | 0,58 | 0,57 | 0,60 | 0,78 |
| Calabria | 0,41 | 1,00 | 0,69 | 0,61 |
| Campania | 0,33 | 0,36 | 0,32 | 0,39 |
| Emilia-Romagna | 0,98 | 1,00 | 0,97 | 0,85 |
| Friuli-Venezia Giulia | 0,92 | 0,88 | 0,92 | 0,91 |
| Lazio | 0,89 | 0,90 | 0,69 | 0,68 |
| Liguria | 0,71 | 0,87 | 0,89 | 0,93 |
| Lombardia | 0,81 | 0,80 | 0,71 | 0,68 |
| Marche | 0,50 | 0,70 | 0,47 | 0,60 |
| Molise | 0,83 | 0,46 | 0,55 | 0,39 |
| Piemonte | 0,79 | 0,79 | 0,83 | 0,84 |
| Puglia | 0,52 | 0,51 | 0,35 | 0,38 |
| Sardegna | 0,46 | 0,46 | 0,50 | 0,34 |
| Sicilia | 0,33 | 0,28 | 0,29 | 0,29 |
| Toscana | 0,58 | 0,55 | 0,56 | 0,58 |
| Umbria | 0,63 | 0,62 | 0,56 | 0,72 |
| Valle d'Aosta | 0,73 | 0,66 | 0,80 | 0,86 |
| Veneto | 0,79 | 0,67 | 0,84 | 0,85 |
| Alto Adige | 0,84 | 0,74 | 0,85 | 0,85 |
| Trentino | 1,00 | 0,89 | 0,87 | 0,96 |

**Legenda:** la scala cromatica utilizzata va dal rosso intenso (efficienza bassa) al verde intenso (efficienza alta).

Guardando all'efficienza nell'erogazione dei servizi distrettuali, la situazione che emerge è ben diversa. Molte delle regioni inefficienti sul piano dei servizi ospedalieri, in questo caso sono ai primi posti durante tutto il periodo di tempo considerato. Ciò è particolarmente vero per Campania e Sicilia. Tra le altre regioni efficienti ci sono Umbria, Abruzzo e Basilicata (Tab. 4.4).
Performance molto negative per Alto Adige, Marche, Piemonte e Friuli Venezia Giulia (Tab. 4.4).

**Tab. 4.4: l'efficienza nell'erogazione dei servizi di assistenza distrettuale, periodo 2010 – 2013, per regione**

| Regione | 2010 | 2011 | 2012 | 2013 |
|---|---|---|---|---|
| Abruzzo | 0,93 | 0,98 | 0,99 | 0,93 |
| Basilicata | 0,96 | 0,86 | 0,97 | 0,93 |
| Calabria | 0,67 | 0,79 | 0,80 | 1,00 |
| Campania | 0,93 | 1,00 | 0,93 | 0,97 |
| Emilia-Romagna | 0,87 | 0,90 | 0,92 | 0,79 |
| Friuli-Venezia Giulia | 0,56 | 0,60 | 0,61 | 0,52 |
| Lazio | 0,78 | 0,87 | 0,51 | 0,49 |
| Liguria | 0,84 | 0,66 | 0,85 | 0,74 |
| Lombardia | 0,63 | 0,62 | 0,61 | 0,58 |
| Marche | 0,63 | 0,52 | 0,46 | 0,51 |
| Molise | 0,68 | 0,83 | 0,83 | 0,76 |
| Piemonte | 0,51 | 0,50 | 0,70 | 0,67 |
| Puglia | 0,84 | 0,87 | 0,89 | 0,83 |
| Sardegna | 0,92 | 0,89 | 0,86 | 0,84 |
| Sicilia | 1,00 | 0,99 | 1,00 | 0,96 |
| Toscana | 0,89 | 0,87 | 0,86 | 0,88 |
| Umbria | 0,96 | 0,94 | 0,93 | 0,93 |
| Valle d'Aosta | 0,79 | 0,76 | 0,78 | 0,74 |
| Veneto | 0,73 | 0,90 | 0,77 | 0,69 |
| Alto Adige | 0,37 | 0,36 | 0,37 | 0,37 |
| Trentino | 0,57 | 0,67 | 0,82 | 0,80 |

**Legenda:** la scala cromatica utilizzata va dal rosso intenso (efficienza bassa) al verde intenso (efficienza alta).

Infine, è possibile valutare l'efficienza complessiva, calcolabile come somma dei due indici parziali appena ottenuti. Oltre ad una somma semplice, è interessante guardare anche ad una somma pesata. I pesi di questa seconda somma sono determinati sulla base delle soglie fissate per il riparto del finanziamento della sanità, che prevede di destinare il 5% alla prevenzione, il 51% all'assistenza distrettuale e 44% all'assistenza ospedaliera (viene escluso il primo elemento, non essendo oggetto di analisi in questo report).

Guardando ai dati sull'efficienza complessiva ottenuta con somma semplice (l'efficienza massima è, qui, pari a 2, mentre la minima resta a 0), sono Emilia Romagna e Trentino (soprattutto negli ultimi due anni) ad evidenziare livelli di efficienza più elevati rispetto al complesso dei servizi di assistenza sanitaria. Buone performance, durante l'intero periodo, anche per Liguria e Veneto (Tab. 4.5).

Male le Marche, la Puglia, le due isole maggiori e l'Alto Adige (Tab. 4.5).

Tra le regioni che, nel tempo, migliorano la propria situazione ci sono Piemonte, Basilicata, Calabria e Valle d'Aosta (Tab. 4.5).

Peggiorano il Lazio e, in maniera meno evidente, la Lombardia (Tab. 4.5).

**Tab. 4.5: l'efficienza nell'erogazione dei servizi sanitari, periodo 2010 – 2013, per regione (somma semplice)**

| Regione | 2010 | 2011 | 2012 | 2013 |
|---|---|---|---|---|
| Abruzzo | 1,48 | 1,50 | 1,55 | 1,47 |
| Basilicata | 1,54 | 1,43 | 1,56 | 1,71 |
| Calabria | 1,08 | 1,79 | 1,49 | 1,61 |
| Campania | 1,26 | 1,36 | 1,24 | 1,35 |
| Emilia-Romagna | 1,85 | 1,90 | 1,88 | 1,64 |
| Friuli-Venezia Giulia | 1,48 | 1,49 | 1,53 | 1,43 |
| Lazio | 1,67 | 1,77 | 1,20 | 1,17 |
| Liguria | 1,55 | 1,52 | 1,74 | 1,67 |
| Lombardia | 1,44 | 1,42 | 1,32 | 1,26 |
| Marche | 1,13 | 1,22 | 0,93 | 1,11 |
| Molise | 1,51 | 1,29 | 1,37 | 1,16 |
| Piemonte | 1,29 | 1,29 | 1,53 | 1,50 |
| Puglia | 1,37 | 1,38 | 1,24 | 1,21 |
| Sardegna | 1,38 | 1,35 | 1,37 | 1,18 |
| Sicilia | 1,33 | 1,27 | 1,29 | 1,25 |
| Toscana | 1,47 | 1,42 | 1,42 | 1,46 |
| Umbria | 1,59 | 1,56 | 1,49 | 1,65 |
| Valle d'Aosta | 1,52 | 1,41 | 1,59 | 1,60 |
| Veneto | 1,52 | 1,58 | 1,61 | 1,54 |
| Alto Adige | 1,21 | 1,10 | 1,22 | 1,22 |
| Trentino | 1,57 | 1,55 | 1,69 | 1,76 |

**Legenda:** la scala cromatica utilizzata va dal rosso intenso (efficienza bassa) al verde intenso (efficienza alta).

Quando si considera l'efficienza complessiva ponderando rispetto al peso che i diversi tipi di assistenza sanitaria dovrebbero avere in termini di quote di spesa dedicate, la situazione resta sostanzialmente simile alla precedente. Migliorano un po' le performance dell'Umbria, nonché quelle delle regioni meridionali che, fatta eccezione per Calabria, Basilicata e, soprattutto, Abruzzo, restano comunque nel gruppo di coda (Tab. 4.6).

**Tab. 4.6: l'efficienza nell'erogazione dei servizi sanitari, periodo 2010 – 2013, per regione (somma pesata)**

| Regione | 2010 | 2011 | 2012 | 2013 |
|---|---|---|---|---|
| Abruzzo | 1,51 | 1,53 | 1,58 | 1,50 |
| Basilicata | 1,56 | 1,45 | 1,59 | 1,72 |
| Calabria | 1,10 | 1,78 | 1,50 | 1,64 |
| Campania | 1,30 | 1,41 | 1,29 | 1,40 |
| Emilia-Romagna | 1,84 | 1,89 | 1,88 | 1,64 |
| Friuli-Venezia Giulia | 1,45 | 1,47 | 1,50 | 1,40 |
| Lazio | 1,67 | 1,77 | 1,19 | 1,16 |
| Liguria | 1,56 | 1,51 | 1,74 | 1,66 |
| Lombardia | 1,43 | 1,40 | 1,31 | 1,25 |
| Marche | 1,14 | 1,21 | 0,93 | 1,10 |
| Molise | 1,50 | 1,32 | 1,39 | 1,18 |
| Piemonte | 1,27 | 1,27 | 1,52 | 1,49 |
| Puglia | 1,39 | 1,40 | 1,28 | 1,24 |
| Sardegna | 1,42 | 1,38 | 1,39 | 1,21 |
| Sicilia | 1,38 | 1,32 | 1,34 | 1,30 |
| Toscana | 1,49 | 1,45 | 1,44 | 1,48 |
| Umbria | 1,61 | 1,59 | 1,52 | 1,66 |
| Valle d'Aosta | 1,53 | 1,42 | 1,58 | 1,59 |
| Veneto | 1,51 | 1,59 | 1,61 | 1,53 |
| Alto Adige | 1,18 | 1,07 | 1,18 | 1,19 |
| Trentino | 1,54 | 1,54 | 1,69 | 1,75 |

**Legenda:** la scala cromatica utilizzata va dal rosso intenso (efficienza bassa) al verde intenso (efficienza alta).

**Conclusioni**

Dalle pagine precedenti emergono alcuni spunti di riflessione:
- le regioni che primeggiano in termini di copertura e qualità dei servizi sono anche quelle più efficienti[12]. Dunque, stante il livello dell'offerta, riescono a garantirlo al costo più basso. Nell'ambito dei servizi ospedalieri ciò è vero per Emilia Romagna, Trentino, Piemonte, Liguria, Friuli Venezia Giulia e Alto Adige; nell'ambito dei servizi distrettuali ciò è vero per Abruzzo, Sardegna ed Emilia Romagna;
- stesso ragionamento, ma rovesciato, può essere fatto per le regioni che hanno efficienza bassa, stante il livello dei servizi erogati. Per quanto riguarda i servizi ospedalieri, questo è vero per Sicilia, Puglia e Campania; per quanto concerne i servizi distrettuali, ciò è vero per l'Alto Adige;
- le regioni con il livello di efficienza complessiva più elevato, quindi Emilia Romagna, Veneto, Trentino e Liguria, sono tutte ubicate al Nord e si caratterizzano per livelli di spesa pro capite sanitaria molto differenziata tra di loro;
- le regioni con livello di efficienza complessiva bassa, quindi Marche, Alto Adige, Sicilia, Puglia e Sardegna, sono distribuite su tutto il territorio nazionale e, anch'esse, non si contraddistinguono per un livello di spesa pro capite particolare. Da sottolineare, però, che più della metà di queste usufruisce dello status di regione a statuto speciale;
- fatte le dovute eccezioni, molte regioni che primeggiano nell'ambito dell'assistenza distrettuale (in particolare quelle meridionali) sono agli ultimi posti quando si guarda all'assistenza ospedaliera, e viceversa (Alto Adige, Piemonte e Friuli Venezia Giulia);
- il sistema sanitario dell'Emilia Romagna è quello che, sia da un punto di vista della copertura del servizio e della sua qualità, sia da un punto di vista dell'efficienza, sopravanza tutti gli altri;
- tra le regioni in piano di rientro, sono Piemonte e Calabria quelle che, nel tempo, riescono a garantire una copertura dei servizi sanitari ad un costo più basso. Molise e Lazio, in particolare, peggiorano la propria efficienza negli anni. Stabile la Campania.

---

[12] In parte, questo effetto è imputabile al tipo di modello utilizzato. Si può, tuttavia, considerare come un elemento di premialità nei confronti di chi riesce a garantire la copertura territoriale migliore dei propri servizi. Per altro, il modello scelto, seppur presenti questo vantaggio nei confronti di chi "offre" di più, è però in grado di dare enfasi non solo agli aspetti quantitativi dei servizi, ma anche a quelli qualitativi, che spesso, in altri lavori, passano in secondo piano.